\documentclass[%
 reprint,
superscriptaddress,
 amsmath,amssymb,
 aps,prl
]{revtex4-1}

\usepackage{graphicx}
\usepackage{dcolumn}
\usepackage[utf8]{inputenc}
\usepackage{bm}
\usepackage[colorlinks=true,linkcolor=blue,citecolor=blue]{hyperref}
\usepackage{float}
\renewcommand{\u}{\bm{u}}
\renewcommand{\b}{\bm{b}}
\renewcommand{\r}{\bm{r}}
\newcommand{\x}{\bm{x}}

\usepackage{amsmath,bm}
\DeclareMathAlphabet\mathbfcal{OMS}{cmsy}{b}{n}

\usepackage[normalem]{ulem}

\usepackage{xcolor}

\newcommand{\figref}[1]{Fig. \ref{#1}}

\begin{document}


\title{Manuscript Title:\\with Forced Linebreak}

\title{Subion Scale Turbulence Driven by Magnetic Reconnection}
\author{D. Manzini}
\email{davide.manzini@lpp.polytechnique.fr}
\affiliation{Laboratoire de Physique des Plasmas (LPP), CNRS, École Polytechnique, Sorbonne Université, Université Paris-Saclay, Observatoire de Paris, 91120 Palaiseau, France}%
\affiliation{Dipartimento di Fisica E.Fermi, University of Pisa, Italy}%

\author{F. Sahraoui}
\affiliation{Laboratoire de Physique des Plasmas (LPP), CNRS, École Polytechnique, Sorbonne Université, Université Paris-Saclay, Observatoire de Paris, 91120 Palaiseau, France}%

\author{F. Califano}
\affiliation{Dipartimento di Fisica E.Fermi, University of Pisa, Italy}%

\date{\today}

\begin{abstract}
\noindent The interplay between plasma turbulence and magnetic reconnection remains an unsettled question in astrophysical and laboratory plasmas. Here we report the first observational evidence that magnetic reconnection drives subion scale turbulence in magnetospheric plasmas by transferring energy to small scales. We employ a spatial \emph{coarse-grained} model of Hall magnetohydrodynamics, enabling us to measure the nonlinear energy transfer rate across scale $\ell$ at position $\bm{x}$. Its application to Magnetospheric Multiscale mission data shows that magnetic reconnection drives intense energy transfer to subion scales. This observational evidence is remarkably supported by the results from Hybrid Vlasov-Maxwell simulations of turbulence to which the coarse-grained model is also applied. These results can potentially answer some open questions on plasma turbulence in planetary environments. \\
\end{abstract}

\maketitle

Decades of observational research have shown that astrophysical plasmas are generally in a turbulent state, the most popular signature being the $ f^{-5/3}$ power-law spectrum of the magnetic field fluctuations over a frequency band often referred to as  the ``inertial range" \citep{coleman68,bavassano82,bruno2013,sahraoui2020}. At higher frequencies (away from the ion transition range  \citep{sahraoui2010}) the spectra steepen to $\sim f^{-2.8}$, which is termed the dispersive or the dissipation range \citep{goldstein94,stawicki2001,sahraoui2009}. The conventional wisdom holds that the inertial range is populated predominantly by incompressible Alfv\'enic fluctuations well described by the classical magnetohydrodynamics (MHD) theory of turbulence \citep{matthaeus82,shebalin83,matthaeus_review,GS95}, while at scales comparable or smaller than the ion characteristic scale, there is an increasing body of evidence that turbulence transitions into  kinetic Alfv\'en waves (KAW) that sustain the energy cascade in the subion range \citep{bale2005,sahraoui2010,chen2013,Salem2012}. {Turbulence can further develop striking features of self-organization in the form of coherent structures such as current sheets (CS), magnetic eddies or magnetic holes, which can be locations of intense energy dissipation and localized particle heating \citep{Yang17,Karimabadi_struct,Wan_struct,Huang_MHoles,Huang_2017_EV,Huang_2018_WHI,CamporealePRL} }{however, it is not clear how, and to what extent those structures contribute to the energy cascade; in particular, a quantitative measure of the associated cross-scale energy transfer is, to date, lacking.}
Here, using in-situ data gathered in the near-Earth space and Hybrid Vlasov-Maxwell simulations {we present the first direct measure of the local cross-scale energy transfer induced by magnetic reconnection (MR) showing that this mechanism can be the main driver of subion scale turbulence providing a net energy transfer to small scales. The importance of this mechanism is backed by} previous numerical works showing that MR can develop on timescales smaller than the nonlinear {(eddy turnover)} time thus being more effective than the fluid-like scale-by-scale interaction in driving subion scale turbulence  \citep{Papini_2019,papini19,franci2017,Cerri2017}.
{Lastly, this new channel for driving subion scale turbulence does not require the presence of an inertial range and as such it} can potentially explain pending observations in the Earth and planetary magnetosheaths. {Data taken from those media showed indeed an ubiquitous $f^{-2.8}$ turbulent spectrum at subion scales regardless of the presence or absence of the inertial range \citep{sahraoui2013,huang2017,sahraoui2020} the latter being characterized by an $f^{-1}$ spectrum and Gaussian statistic of the magnetic field increments at the large (MHD) scales \citep{hadid2015,tao15,huang2017,huang2020}}. This is somehow conflicting with the standard Kolmogorov picture of scale-by-scale (fluid) cascade from the inertial to the dissipation range. 

{\textit{The coarse graining approach.}--}MR in magnetized plasmas is a process that occurs in very {thin CS \citep{sundkvist2007,biskamp_magnetic_1996,shay2001,birn_geospace_2001,mandt_transition_1994,burch2016,cassak2007} and the importance of its interplay with plasma turbulence is testified by the wealth of literature on the subject \citep{Lazarian_rec,matthaeus_turbulent_1998,Beresnyak_2017,loureiro_role_2017,boldyrev_role_2019,Kowal_2017,Adhikari,servidio_prl}. In this work we aim at quantifying the {\it local} (in space) cross-scale energy transfer in reconnecting CS. To do so, it is required to go beyond the popular Kolmogorov $4/5$-law, which provides the turbulent transfer rate at scale $\ell$ after averaging over all spatial positions ${\x}$ \citep{politano1998,galtier2008,banerjee2014,hellinger2018,andres2018,simon2021,simon2022}}. To this end we employ the spatial {\em coarse-graining} (CG) approach  \citep{eyink_locality_2005,aluie_compressible_2011,aluie_coarse-grained_2017,Yang17,Yang_2022,CamporealePRL,Cerri_camporeale} and, in particular, the model for incompressible Hall-MHD \citep{manzini2022}. 
Here the CG fields are defined by $\bar{f}_\ell(\x)=\int d\r G_\ell(\r)f(\x+\r)$ where $G_\ell$ is a centered, normalized, filtering function with variance of order $\ell^2$. The filtered quantities $\bar{\u}_\ell, \bar{\b}_\ell$ retain only scales  $\geq \ell$ and are used to derive the equations for the large and small scale energy densities \citep{manzini2022}. Of particular importance is the quantity $\pi_\ell(\x)$ describing the energy transfer rate across scale $\ell$ at position $\x$. In Alfv\'en units it reads:
\begin{equation}
    \pi_\ell(\x)=-\rho_0\nabla\bar{\u}_\ell:\bm{\tau}_\ell-\bar{\bm{j}}_\ell\cdot\mathbfcal{E}_\ell
    \label{eq:pi_def}
\end{equation}
where $\rho_{_0}$ is the mean mass density, $\u=(m_i\bm{v_i}+m_e\bm{v_e})/(m_i+m_e)$ the MHD velocity, $\bm{b}=\bm{B}/\sqrt{\mu_0\rho_{_0}}$ the local Alfv\'en speed and $\bm{j}=\nabla\times\bm{b}$ the normalized current density. Furthermore, $\bm{\tau}_\ell$ and $\mathbfcal{E}_\ell$ (the sub-scale electric field) are second order tensors that write :
\begin{equation*} 
\begin{split}
    \bm{\tau}_\ell&=\overline{(\u\u)}_\ell-\bar{\u}_\ell\bar{\u}_\ell-\left[\overline{(\bm{b}\bm{b})}_\ell-\bar{\bm{b}}_{\ell}\bar{\bm{b}}_{\ell}\right]\\
    \mathbfcal{E}_\ell=&\overline{(\u\times\bm{b})_\ell}-\bar{\u}_\ell\times\bar{\bm{b}}_\ell-d_i\left[\overline{(\bm{j}\times\bm{b})_\ell}-\bar{\bm{j}}_\ell\times\bar{\bm{b}}_\ell\right]
    \end{split}
\end{equation*}
$d_i$ being the ion inertial length. At scale $\ell$ and position $\bm{x}$, nonlinearities transfer part of the large scale energy density $\bar{E}_\ell=\rho_0(|\bar{\u}_\ell|^2+|\bar{\b}_\ell|^2)/2$ to smaller scales at a rate $\pi_\ell(\x)$ . This can easily be inferred from the CG energy equation
\begin{equation}
\partial_t\bar{E}_\ell+\nabla\cdot\bar{\mathbfcal{J}_\ell}=\mathcal{D_\ell}+\mathcal{F_\ell}-\pi_\ell,
\end{equation}
where $\mathcal{D_\ell}, \mathcal{F_\ell}$ denote the large scale effects of dissipation and forcing and $\nabla\cdot\bar{\mathbfcal{J}_\ell}$ encloses all the terms driving spatial transport of the large scale energy $\bar{E}_\ell$ (see \citep{manzini2022} and references therein for details).
\begin{figure}
    \centering
    \includegraphics[width=0.5\textwidth]{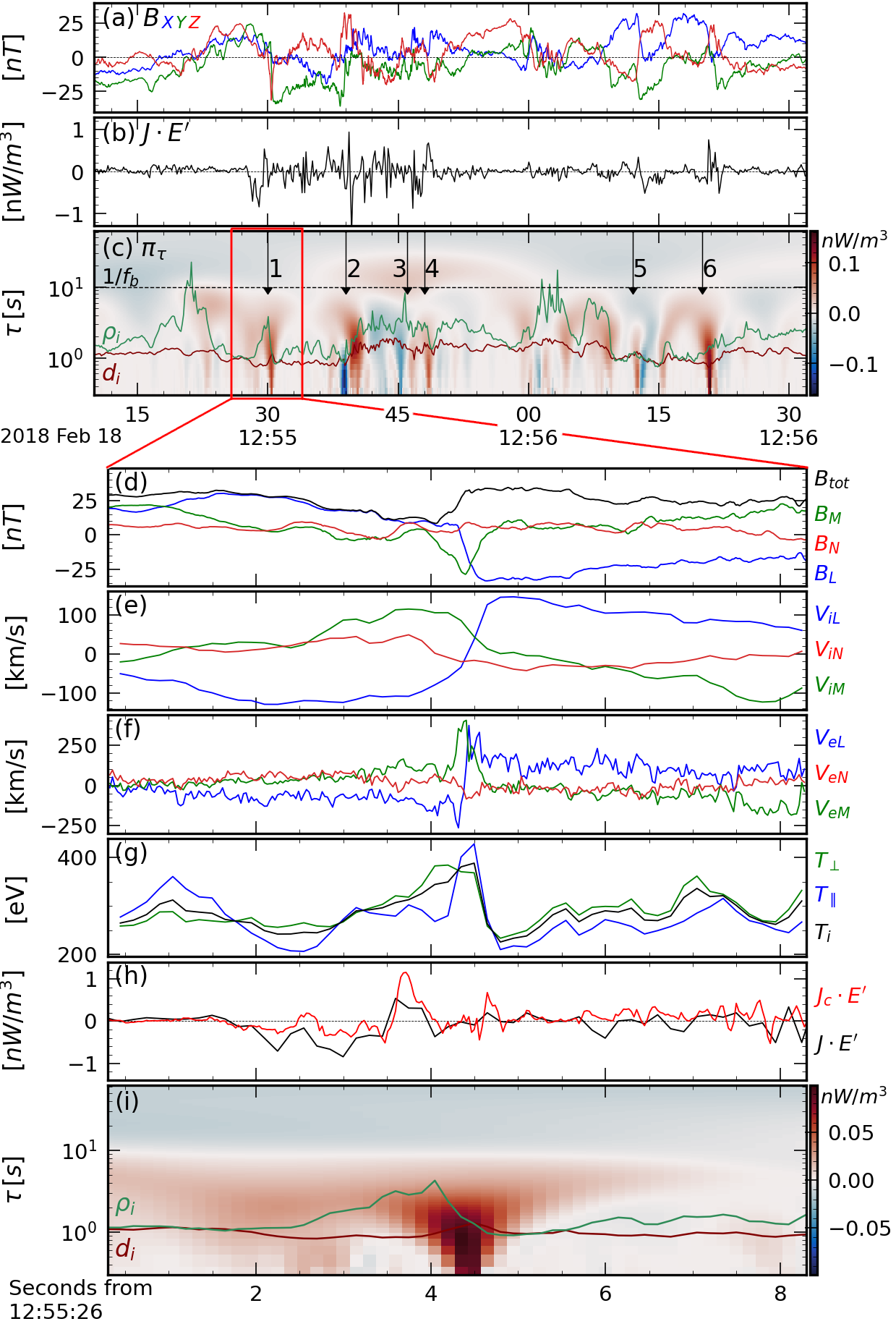}
    \caption{MMS1 data: (a) the three components of the magnetic field in GSE (Geocentric Solar Ecliptic); (b) $\bm{J}\cdot\bm{E}'$, where $\bm{E}'$ is the electric field in the electron rest frame; (c) the estimated transfer rate $\pi_\tau(t)$ using the CG Hall-MHD model. The curves in panel (c) indicate the local Taylor shifted ion Larmor radius $\rho_i$, inertial length $d_i$, and the inverse of the frequency of the spectral break $1/f_b$ denoted in Fig. \ref{fig:MF_Spectrum}. {Arrows 1-6 show reconnecting CS.} Panels (d-i) are a zoom on a reconnecting CS (the vector components are given in LMN coordinates). In panels (e) and (f) the background mean flow $\langle v_i\rangle$ was removed to highlight the super-Alfv\'enic ($V_{A}=B_0/\sqrt{\mu_0\rho_0}=120 km/s$) outflows in the $L$ direction. In panel (h) $\bm{J}_c=\nabla\times\bm{B}/\mu_0$ (red) is the electric current computed using the curlometer technique plotted for comparison.}
\label{fig:my_foreshock crossing} 
\end{figure}{When applying the CG technique to the time series measured onboard spacecraft, the CG operation at a given temporal resolution $\tau$ is defined as}
{ $\bar{f}_\tau(t)=\int dt' G_\tau(t')f(t+t')$. The time scale $\tau$ can then be converted into a spatial scale $\ell$ assuming the Taylor hypothesis, i.e. $\ell\sim \langle u\rangle \tau$, where $\langle u\rangle$ is the mean flow velocity.}

{\textit{Data selection and analysis.}--}{We use data from the Magnetospheric Multiscale (MMS) Mission that enable us to compute the velocity strain in equation \eqref{eq:pi_def} using the gradiometer technique \citep{curlometer}.} The magnetic and electric field data come respectively from the FluxGate Magnetometers (FGM) \cite{russell_magnetospheric_2016} and the Electric Double Probe (EDP) \cite{lindqvist_spin-plane_2016,ergun_axial_2016}. The plasma data were measured by the Dual Ions and Electrons Spectrometer instrument (DIS, DES) \cite{pollock_fast_2016}, and were used to compute the electric current $\bm{J}=n_ee(\bm{v}_i-\bm{v}_e)$. We consider only the electron density under the assumption of quasi-neutrality. When combining MMS's products of different temporal resolutions, we first low-pass filter to the  frequency of the least resolved quantity involved in the computation and then we interpolate to common epochs.
 {Furthermore,} since the CG model we use is derived for incompressible Hall-MHD, {in the following analysis} we select data intervals that feature dominant incompressible Alfv\'enic fluctuations as evidenced by their weak magnetic compressibility $C_\parallel=\langle\delta B_\parallel^2/\delta B^2\rangle\lesssim0.2$ when averaged over the frequency range $[0.01,1]\,$ Hz.

The first data set used in this work was measured by MMS1 on day 2018-02-18 as it was travelling in the terrestrial magnetosheath after several magnetopause crossings about 11:20 (not shown). The selected observables are shown in Fig.\ref {fig:my_foreshock crossing}. 
\begin{figure}
    \centering
    \includegraphics[width=0.5\textwidth]{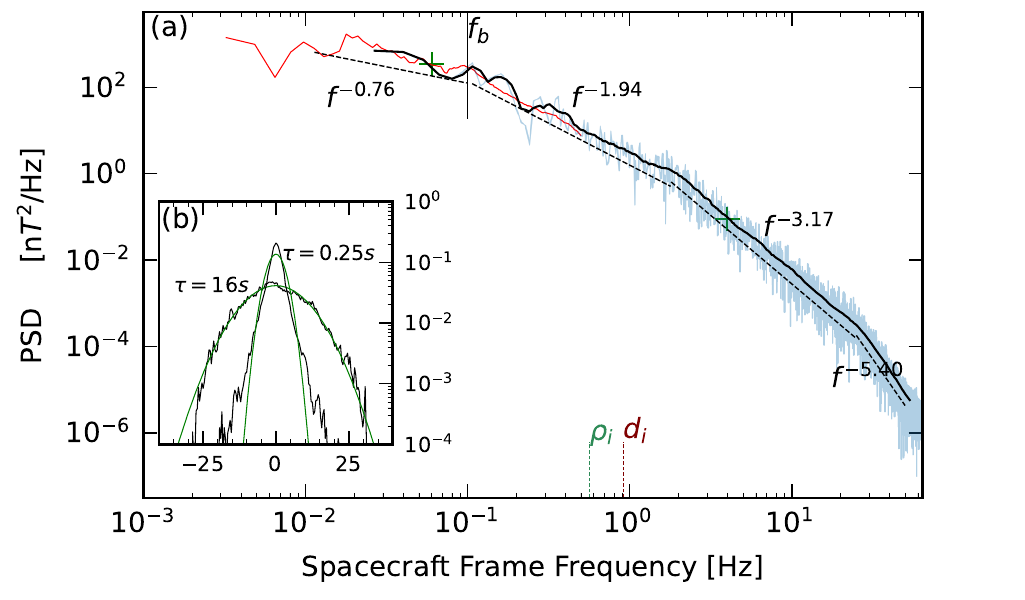}
    \caption{Magnetic field power spectrum for the interval displayed in \figref{fig:my_foreshock crossing} (a), dotted lines denote power-law fits within the different spectral ranges. The spectrum at low frequency (red curve) is computed over the full burst mode interval, 12:52:43-13:03:03. Panel (b) shows the histogram of $\delta B(\tau)=|\bm{B}|(t+\tau)-|\bm{B}|(t)$ for two values of $\tau$ marked in (a) with a green cross. Superimposed is a Gaussian with matching first and second order moments (green curves).}
    \label{fig:MF_Spectrum}
\end{figure}
{The magnetic field in }panel (a) shows sharp gradients resulting in many intense CS where strong energy transfers occur between the electromagnetic fields and the plasma particles as attested by the quantity $\bm{J\cdot\bm{E}'}=\bm{J}\cdot(\bm{E}+\bm{v}_e\times\bm{B})$ shown in \figref{fig:my_foreshock crossing}(b). The $\pi_\tau(t)$ scalogram in panel (c) measures, at each time $t$, the energy transfer sustained by non-linearities across scale $\tau$ . A closer analysis of $\pi_\tau(t)$ in \figref{fig:my_foreshock crossing} (c) reveals that most structures associated with intense energy transfers correspond to locations of ongoing MR (denoted with an arrow). An event is deemed to be a reconnecting CS (RCS) when, after rotating the data in the minimum variance frame (LMN), a $B_L$ reversal together with ion and electron outflows $|v_{i,L}|\gtrsim 0.5V_A, \,|v_{e,L}|\gtrsim V_A$ are observed, $V_A=B_0/\sqrt{\mu_0\rho_0}$ being the average Alfv\'en speed. For each RCS (at a given $t^\star$) the behaviour of $\pi_\tau(t^\star)$ emphasizes scales where the nonlinear transfer associated with MR becomes effective. Interestingly, we observe that even if a weak energy transfer is initiated at scale $\tau\sim 10\,s$, it intensifies at scales comparable with the Taylor shifted ion scales  $d_i, \rho_i$, with a better agreement with the latter. These results are clearly seen in \figref{fig:my_foreshock crossing} (d-i), which features a zoom on a single reconnecting CS together with some trademarks of MR, \emph{e.g.} the magnetic field reversal in the minimum variance frame and super-Alfv\'enic $L$-directed outflows. Notice furthermore that the intensification of the transfer rate $\pi_\tau(t)$ at subion scales correlates well with a local increase in ion parallel and perpendicular temperatures and in $\bm{J}\cdot\bm{E}'$. The other RCS were also found (not shown) to energize locally ions and/or electrons, in the form of heating and/or acceleration (i.e., jets). 
The weakening of the energy transfer at large scales ($\tau>10 \,s$) in \figref{fig:my_foreshock crossing}(c) is consistent with the magnetic field spectrum in \figref{fig:MF_Spectrum} which flattens to $\sim f^{-0.8}$ at those scales and with the Gaussian distribution of the field increments at $\tau=16$s (\figref{fig:MF_Spectrum}(b)). In contrast, the very intense and mostly positive energy transfers to smaller scales are reflected in the power-law spectrum observed at high frequency ($f\gtrsim 2\,$Hz) and the departure from Gaussianity of the increments histogram computed for  $\tau=0.25\,s$ (\figref{fig:MF_Spectrum}(b)). Note that the onset of the (weak) transfer at large scales ($\tau\sim 10\,s$) corresponds roughly to the inverse of the frequency of the first spectral break denoted in \figref{fig:MF_Spectrum}(a) that marks the transition into the Kolmogorov-like inertial range.

\begin{figure}
    \centering
    \includegraphics[width=0.5\textwidth]{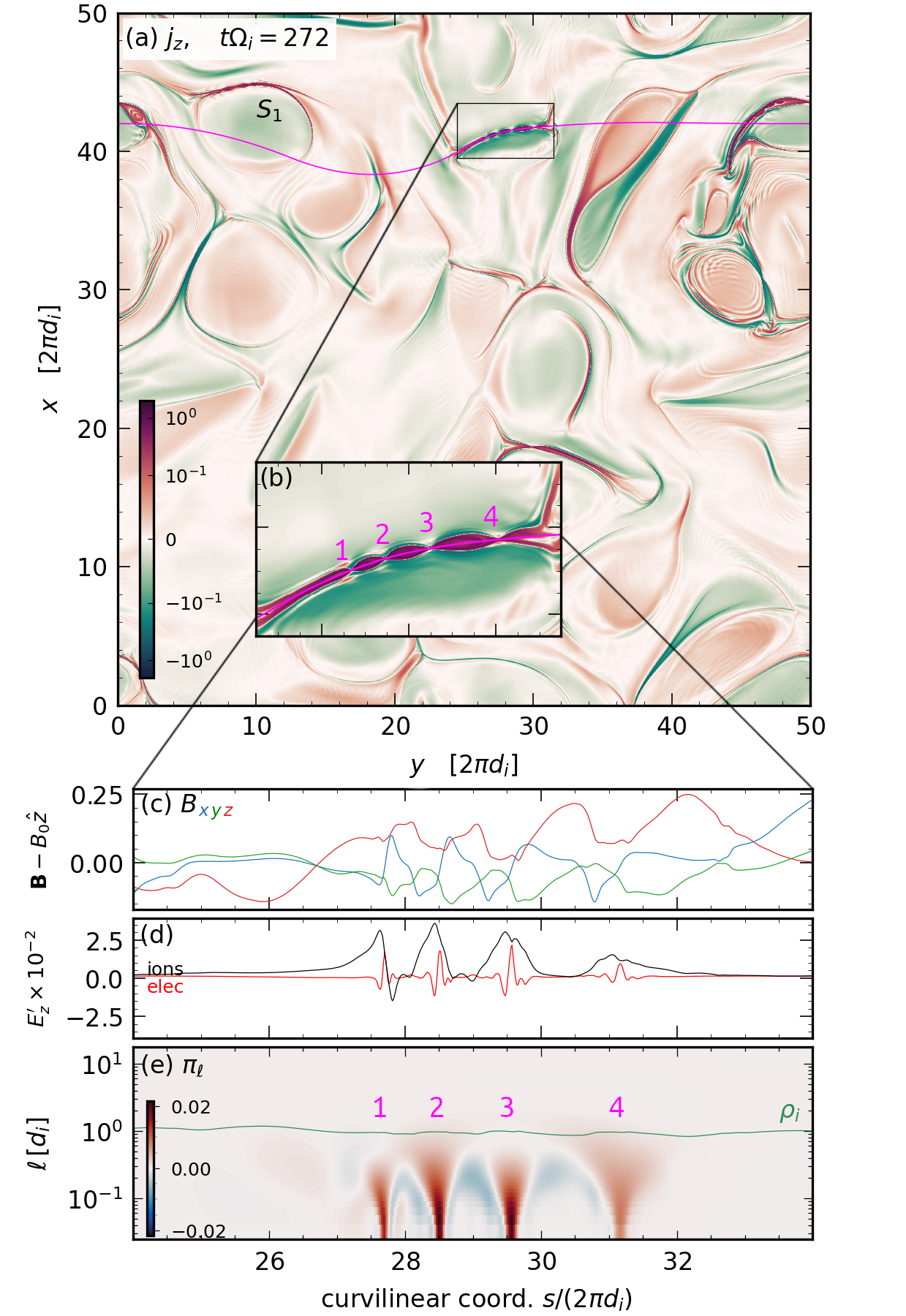}
    \caption{Panel (a) shows the out-of-plane current $j_z$ at $t\Omega_i=272$ {and the trajectory $S_1$ crossing a reconnecting region highlighted with a black frame and in the inset (b). Panels (c-e) show  quantities interpolated along $S_1$ in the reconnecting region:} (c) the three components of the magnetic field (the background field $B_0\hat{z}$ was removed), (d) the out-of-plane electric field in the ion (electron) rest frame as a marker of the ion (electron) diffusion region, (e) the nonlinear transfer rate $\pi_\ell(\x)$ as a function of scale $\ell$ and position along the curve. {Numbers 1-4 denote the reconnection sites. }}
    \label{fig:my_HVM_cut}
\end{figure}

{\textit{Support from numerical simulations.}--}To gain further insight into the interplay between MR and the associated localized cross-scale energy transfer we study the emergence of MR in a turbulence simulation performed using a Hybrid Vlasov-Maxwell (HVM) code \citep{valentini2007} where ions evolve following the Vlasov equation and electrons are described as a fluid and respond through the generalized Ohm’s law including their inertia. In this approach we assume quasi-neutrality and an electron isothermal equation of state. We consider an initial unity proton and electron plasma beta (ratio between the thermal and magnetic pressure of each species), i.e. $\beta_i=\beta_e=1$, impose a homogeneous out-of-plane magnetic field and a mass ratio $m_i/m_e=100$. The 2D-3V HVM equations (2D in space, 3D in velocity) are solved on a squared $N_x\times N_y$ grid of $3072^2$ points spanning a real-space domain of size $L_x = L_y = 50\times 2\pi d_i$. The ion velocity space is limited to $\pm 5v_{th,i}$ and sampled on a cubic grid of $51^3$ points.
The simulation is initiated by imposing random, isotropic magnetic fluctuations with corresponding wave numbers lying in the  range $0.02 \leq |\bm{k}|d_i\leq 0.12$ (the largest admitted by the box). At $t\Omega_i\sim 240$, before turbulence fully develops, many CS grow unstable and start to reconnect (\figref{fig:my_HVM_cut}(a)).
{\figref{fig:my_HVM_cut}(c-e) display quantities interpolated in the reconnecting regions along the 1D curve  $S_1$ crossing  the simulation domain shown in panel (a). The electron and} ion diffusion regions {(determined by non-zero $E^{\prime,i/e}_z$ in \figref{fig:my_HVM_cut}(d)), show enhanced and positive $\pi_\ell(\x)$, a signature of the ongoing generation of small scale fluctuations}. In remarkable agreement with MMS data, we observe that the nonlinear transfer associated with MR becomes effective at the scale $\ell\sim d_i\sim \rho_i$ with little to no energy transfer at larger scales. 
\begin{figure}
    \centering
    \includegraphics[width=0.5\textwidth]{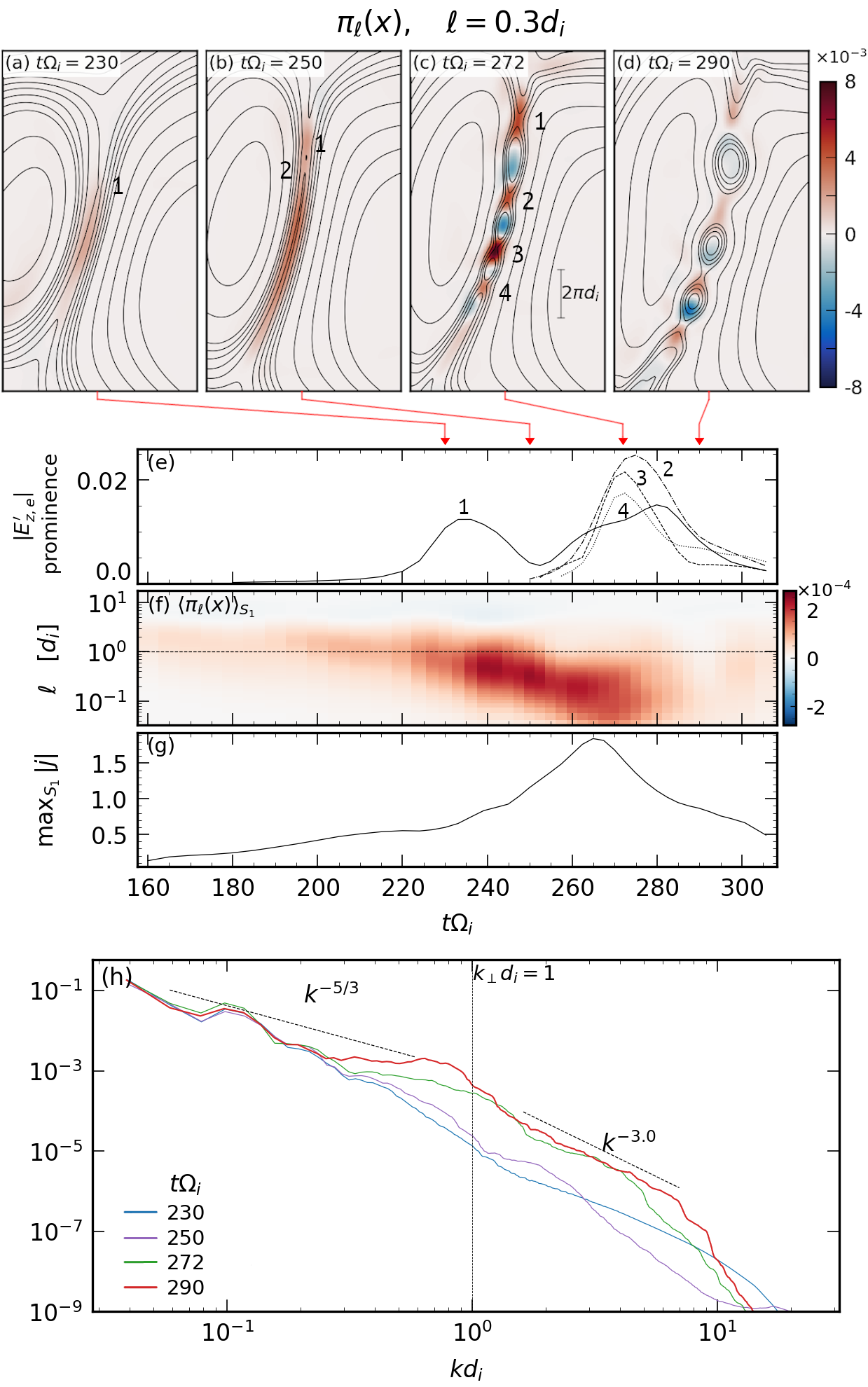}
    \caption{ {Panels (a-d) show the cross-scale transfer rate $\pi_\ell(\x)$ computed at $\ell=0.3d_i$ within the current sheet undergoing MR. Superimposed are the magnetic field lines. 
    Panel (e) shows the prominence of the $E^{\prime,e}_z$ peak at each X-point (denoted 1-4) and referenced in panels (a-d) as a function of time while in panels (f-g) we display $\pi_\ell$ averaged along the trajectory $S_1$  and the max value of $|j|$ along the same trajectory. Lastly Panel (h) shows the evolutuion of the magnetic field spectrum computed along $S_1$}.}
    \label{fig:my_time_analysis}
\end{figure}
{To strengthen the evidence that MR is the driver of the subion scale energy transfer, we follow its time evolution using as tracer of its development the prominence (or \emph{relative height}) of the $|E^{\prime,e}_z|$ peak  at each X-point and the max of $|j|$ computed along $S_1$. Notice that since the reconnecting structure is not fixed in space, the trajectory is slightly displaced at each time-step to follow the features of interest. Alongside these two quantities we compute the average of $\pi_\ell(x)$ along $S_1$ to monitor the net energy transfer to small scales, these quantities are displayed in \figref{fig:my_time_analysis} (e-g). We also selected four timesteps, denoted with red markers in panel (e), for which we show the magnetic field lines together with the quantity $\pi_\ell(\x)$ at $\ell=0.3 d_i$ in the reconnecting region (panels (a-d)) and the magnetic field spectrum computed along $S_1$ in panel (h).\\
At early times only one X-point is observed (denoted 1 in panels (a-e) ) and the $|E^{\prime,e}_z|$ peak prominence grows in time up to $t\Omega_i\sim 235$. Simultaneously, we observe a strong enhancement of $\langle\pi_\ell(x)\rangle_{S_1}$, sign that nonlinear interactions are starting to transfer energy to smaller scales effectively. Notice that at this time the energy transfer is most effective across $\ell\sim d_i$ and not at smaller scales; however as time advances, energy moves to smaller and smaller scales and energy transfer across those small scales becomes effective as well.}
{During this phase the max of $|j|=|\nabla\times\bm{B}|$ increases monotonically, signature of the progressive building of small-scale magnetic fluctuations}. {Around $t\Omega_i\sim 260$ the CS develops three more reconnecting sites (denoted 2-4) leading to an increase in the corresponding $|E^{\prime, e}_z|$ peak prominence. While locally there is intense cross scale transfer (\figref{fig:my_time_analysis} (c)) the presence of magnetic islands is associated with a complex and fully nonlinear dynamics that transfers energy in both directions, from large to small scales and vice-versa ($\pi_\ell(\x)<0$)}. {This mechanisms weaken the averaged cascade rate and eventually (after $t\Omega_i\sim 270$) nonlinearities { are unable to sustain the growth of} the max of $|j|$ which starts to decline in the reconnecting region.
This picture is further confirmed by analyzing the magnetic field spectrum in \figref{fig:my_time_analysis} (h) computed along the chosen trajectory. As MR proceeds, even if the MHD scales are still far from the $k^{-5/3}$ scaling, the reconnecting region develops a power-law spectrum with a slope $\sim k^{-3}$ similar to those reported from spacecraft observations in various regions of the near-Earth space \citep{sahraoui2009,sahraoui2013,huang14} and from numerical simulations \citep{howes2011}. {Note that in other regions with no reconnecting CS (not shown), no clear power-law develops at the same times and a turbulent regime is reached much later in the simulation}. These results clearly demonstrate that MR generates {\it locally} small scale magnetic fluctuations that populate the subion range spectrum. Both in the simulation and in the MMS data, the background plasma lies generally away from the (linear) instability thresholds of the three main known ion-scale instabilities (oblique firehose, cyclotron and mirror), thus ruling out this mechanism as a potential driver of subion scale turbulence \citep{sahraoui2006}.

{\textit{Conclusions}.--}In this Letter we employ a spatial coarse-grained model of incompressible Hall-MHD and demonstrate that magnetic reconnection drives subion scale turbulence in the MMS data taken in the turbulent terrestrial magnetosheath. The results are consistently confirmed using the same model on the {2D HVM} simulation featuring magnetic reconnection: the ion and the electron diffusion regions experience an intense and positive energy transfer to small scales, in agreement with the formation of a power-law spectrum at subion scales while no cascade is observed at larger scales. These results open new pathways to investigate the interplay between turbulence, reconnection and energy dissipation in collisionless magnetized plasmas and can potentially explain spacecraft observations in various planetary magnetosheaths that showed the ubiquity of  turbulent fluctuations at subion scales even when no energy cascade emerges from the larger (MHD) scales. {The 2D numerical simulation employed may not reflect realistic properties of 3D turbulence and magnetic reconnection. However, we do not expect that the qualitative picture that emerges from the numerical study will change significantly in 3D. A detailed study of the latter case, including the effect of the reduction of dimensions will be tackled in future works.} 
\begin{acknowledgements}
\noindent FS acknowledges useful discussions with S. Huang, DM thanks L. Sorriso-Valvo and G. Cozzani for their precious comments. We acknowledge the Italian supercomputing center CINECA where the simulations have been performed under ISCRA grant.  MMS data come from CDPP/AMDA \cite{AMDA} and NASA GSFC's Space Physics Data Facility's CDAWeb. The python client SPEASY \cite{Speasy} was used for data retrieval.
\end{acknowledgements}

\bibliographystyle{apsrev4-1} 

%

\end{document}